\documentclass[10pt,letterpapers,floats]{article}
\usepackage{amsfonts}
\usepackage{multirow}
\usepackage[multiple]{footmisc}
\usepackage{booktabs}
\newcommand{\nc}{\newcommand}

\renewcommand{\thefootnote}{\fnsymbol{footnote}}
\nc{\fig}[5]{ 
\begin{figure}[!htbp]
    \begin{center}
    \leavevmode
    \centerline{
        \includegraphics[width=#1, height=#2]{#3}
        }
    \caption[]{#4}
    \label{#5}
    \end{center}
\end{figure}}
\nc{\figs}[8]{
\begin{figure}[!htbp]
    \begin{center}
    \leavevmode
    \centerline{
        \includegraphics[width=#1, height=#2]{#3}
        \includegraphics[width=#4, height=#5]{#6}
         }
    \caption[]{#7}
    \label{#8}
    \end{center}
\end{figure}}

\begin{document}
\begin{flushright}
{\rm hep-th/0702170}
\end{flushright}
\vspace{12mm}
\begin{center}
{{{\Large {\bf Entropy Function and Universal Entropy of Two-Dimensional Extremal Black Holes}}}}\\[10mm]
{Seungjoon Hyun$^{a,c}$\footnote{email:hyun@phya.yonsei.ac.kr},
Wontae Kim$^{b,c}$\footnote{email:wtkim@sogang.ac.kr}, 
John J. Oh$^{a}$\footnote{email:john5@yonsei.ac.kr}, and
Edwin J. Son$^{a}$\footnote{email:ejson@yonsei.ac.kr}}\\[8mm]

{{${}^{a}$ Institute of Physics and Applied Physics, Yonsei University, Seoul, 120-749, Korea\\[0pt]       
${}^{b}$ Department of Physics, Sogang University, Seoul, 121-742, Korea\\[0pt]
${}^{c}$ Center for Quantum Space-time, Sogang University, Seoul, 121-742, Korea}\\[0pt]
}
\end{center}
\vspace{2mm}
\begin{abstract}
The entropy for two-dimensional black holes is obtained through the entropy function with
the condition that the geometry approaches an $AdS_2$ spacetime in the near horizon
limit. It is shown that the entropy is universal and proportional to the value of the dilaton
field at the event horizon as expected. We find this universal
behavior holds even after the inclusion of higher derivative terms,
only modifying the proportional constant. More specifically, a variety
of models of the dilaton gravity in two dimensions are considered, in
which it is shown that the universal entropy coincides with the
well-known results in the previous literatures.
\end{abstract}
\vspace{5mm}

{\footnotesize ~~~~PACS numbers: 04.70.Dy, 04.60.Kz}


\vspace{1.5cm}

\hspace{11.5cm}{Typeset Using \LaTeX}
\newpage
\renewcommand{\thefootnote}{\arabic{footnote}}
\setcounter{footnote}{0}

\section{Introduction}

The well-known area law for the black hole entropy that the entropy of black holes is proportional
to the area at the horizon might be questionable in two dimensions
since the entropy of two-dimensional black hole seems to be zero. In
other words, the area of two-dimensional black holes simply vanishes
since the hypersurface of the one-dimensional space is a
point. Indeed, this is closely related to the fact that the theory of
gravity in two dimensions is trivial in the Einstein frame.
However, in two-dimensional dilaton gravities~\cite{cghs,jt}, it has
been shown that the entropy is proportional to the value of the dilaton
field at the horizon from the thermodynamic relation and the
statistical count of microstates~\cite{np,kr,cm,cs,klp:thermo}. This
is supported by the area law in higher-dimensional theory of gravity
since the dilaton field is associated with the radius of the
compactified coordinate, provided we consider the two-dimensional
dilaton gravity as the dimensional reduction from the
higher-dimensional theories.

On the other hand, it was shown that the first law of the black
hole thermodynamics holds in any theory of gravity derived from the
Hamilton-Jacobi's analysis~\cite{sw}, and the generalized definition
of the entropy that is simply $2\pi$ times the Noether charge at
the horizon of the horizon Killing field was suggested~\cite{wald}.
This Noether charge method is useful in studying dynamical black holes and
higher derivative theories of gravity since one can no longer use
the area formula in those theories. Then, it is plausible to apply this method to
two-dimensional black holes, which do not obey the area law.
It, however, is not simple to evaluate the entropy from the general
form of the two-dimensional dilaton gravity by use of the method due
to its complexity.

Recently, it has been conjectured that 
the macroscopic entropy agrees to the weak-coupling microscopic
entropy as long as a certain condition holds, namely, that the geometry
near the horizon is given by $AdS_2 \times S^{D-2}$, for all extremal
black holes~\cite{sen}. The entropy function derived from Wald's
formula is defined by, firstly, integrating the Lagrangian density
over $S^{D-2}$, and then taking the Legendre transform of the
resulting function with respect to the electric fields with
multiplication of an overall factor $2\pi$.
The near horizon geometry of the extremal black holes is determined by
extremizing the entropy function and the black hole entropy is given
by the extremum value of the entropy function. This general method
offers an easier way to calculate the black hole entropy in the most
of extremal charged black holes.

In this paper, we  apply this method to the general form of the two-dimensional
scalar-tensor gravity theory, assuming that the near horizon geometry
approaches $AdS_2$. We show that the (extremal) black hole entropy has
a universal form proportional to the value of the dilaton field at the
horizon. We also find that the model, after the inclusion of higher
derivative terms, admit the solution which has $AdS_2$ geometry with
$SO(2,1)$ isometry. In this case, therefore, we find that the
universal behavior holds even after the inclusion of higher derivative
terms, modifying only the overall proportionality constant.

In section~\ref{sec:universal}, we obtain the universal behavior of the extremal black
hole entropy for the generic two-dimensional scalar-tensor gravity
with the gauge field. We also consider the possible higher-derivative
corrections in the action and find their implications in the black
hole entropy. In section~\ref{sec:cghs}, we apply the method to the two-dimensional
gravity which is a low energy effective theory of the heterotic
strings.
In section~\ref{sec:jt}, we consider the Jackiw-Teitelboim model with the extra
gauge field. In section~\ref{sec:dc}, we obtain the black hole entropy in
two-dimensional doubly charged CGHS model.
In section~\ref{sec:discussion}, we draw some discussions.

\section{Universal entropy in two-dimensional extremal black holes}
\label{sec:universal}

We now calculate the entropy function for two-dimensional
scalar-tensor gravity.
A general form of the two-dimensional scalar-tensor gravity action coupled
to a gauge field strength $F_{\mu\nu}$ is given by
\begin{equation}
  I = \frac{1}{2\pi} \int d^2 x \sqrt{-g} \Phi \left[
    R + T(\Phi) (\nabla \Phi)^2 + \lambda^2 U(\Phi)- \frac14 V(\Phi) F^2
    \right], \label{action}
\end{equation}
where $T(\Phi)$, $U(\Phi)$, and $V(\Phi)$ are arbitrary
functions with respect to a scalar field $\Phi$, and $\Lambda$ is a 
two-dimensional cosmological constant. In many cases this
two-dimensional theory comes from the dimensional reduction of the
higher-dimensional theories. The theory may also admit charged black
hole solutions, including extremal black hole solutions. We would like
to find out the general form of the entropy of these extremal black
holes following the procedure in ref.~\cite{sen}. It is, therefore,
natural to assume that the extremal black hole solutions reduces
$AdS_2$ in the near horizon limit, with $SO(2,1)$ symmetry. Then, the
near horizon solution of the extremal black hole with charge $q$ can
be, generically, written in the form of
\begin{eqnarray}
  ds^2 &=& v \left( -r^2 dt^2 + \frac{dr^2}{r^2} \right),
  \nonumber \\
  \Phi &=& u,  \label{sols:NH} \\
  F_{rt} &=& \varepsilon, \nonumber
\end{eqnarray}
where $v$, $u$, and $\varepsilon$ are constants to be determined in
terms of the charge $q$ and the cosmological constant $\Lambda$. Note
that the covariant derivatives of the Riemann tensor, the scalar
field, and the gauge field strength all vanish in this near horizon
geometry, which plays an important role to construct Sen's entropy
function from Wald's entropy formula.
For these field configurations, the Lagrangian density becomes a function of $v$, 
$u$, and $\varepsilon$;
\begin{eqnarray}
  f(v,u,\varepsilon) &=& \frac{1}{2\pi}
  \sqrt{-g} \Phi \left[ R + T(\Phi) (\nabla
  \Phi)^2 + \lambda^2 U(\Phi) - \frac14 V(\Phi) F^2 \right] \nonumber \\
  &=& \frac{v u }{2\pi}\left[ -\frac2v + \lambda^2 U(u) + \frac{V(u)}{2v^2}
  \varepsilon^2 \right],  \label{L:density}
\end{eqnarray}
and from which  one finds the electric charge as
\begin{equation}
  q = \frac{\partial f}{\partial \varepsilon} = \frac{uV(u)}{2\pi v}
  \varepsilon.  \label{charge}
\end{equation}

Now, the entropy function~\cite{sen} is defined as the Legendre transform of
the Lagrangian density with respect to the gauge field $\varepsilon$,
\begin{eqnarray}
  F(v,u,q) &=& 2\pi [ q\varepsilon - f(v,u,\varepsilon)] \nonumber \\
  &=& v u \left[ \frac2v - \lambda^2 U(u) +
  \frac{2\pi^2q^2}{u^2V(u)} \right].
  \label{ent:fnc}
\end{eqnarray}
The undetermined parameter $u, v$ can be fixed by the equations of
motion, which becomes the extremum equations as 
\begin{eqnarray}
  \left. \frac{\partial F}{\partial v} \right|_{(u_e,v_e)} &=& u_e \left[ -\lambda^2 U(u_e) +
    \frac{2\pi^2q^2}{u_e^2V(u_e)} \right] = 0,  \label{v:eqn} \\
  \left. \frac{\partial F}{\partial u} \right|_{(u_e,v_e)}  &=& \left[ 2 - \lambda^2 v_e
    \left( U(u_e) + u_e U'(u_e) \right) - \frac{2\pi^2v_eq^2}{u_e^2V^2(u_e)} \left(
    V(u_e) + u_e V'(u_e) \right) \right]
    = 0,  \label{u:eqn}
\end{eqnarray}
where the prime ($\prime$) denotes the derivative with respect to $u$.  The
entropy is given by the value of the entropy function at the extremum,
\begin{equation}
  S_\mathrm{BH}(q) = F(v_e,u_e,q) = 2 u_e  \label{entropy}
\end{equation}
from eqs.~(\ref{ent:fnc}) and (\ref{v:eqn}).
Note that the entropy is proportional to the overall `effective' coupling constant  associated
with the dilaton field, $u$, as is expected. The effective Newton
constant in two dimensions is given by $\kappa^2={2\pi}/{\langle\Phi\rangle}$
where $\langle\Phi\rangle$ is an expectation value of the dilaton field. In this
set up, it is natural to take the value of the dilaton field at the
horizon, and hence the effective two-dimensional Newton constant
becomes $ \kappa^2={2\pi}/{u_e}$.
In this sense, this is a
universal result in that it depends only on the overall coupling constant
of the Lagrangian, irrespective of the specific form of the
action.\footnote{We find the overall factor difference in black hole
  entropy between our results and the results in the literatures by
  $2\pi$ or $\pi$. This is purely due to the convention of
  two-dimensional Newton constant.} The alternative form of the black
hole entropy, which will be useful later on, is given by
\begin{equation}
S_\mathrm{BH}(q) =  \left. \frac{2\pi^2vq^2}{V}[\ln(u^2UV)]'  \right|_{(u_e,v_e)} 
\end{equation}

Now let us consider the effect of higher derivative terms. Since in
two dimensions Riemann tensor and Ricci tensor can be expressed in
terms of Ricci scalar, it is sufficient to consider the  higher
derivative terms of the form $R^n$. If one includes the generic higher
derivative terms with additional terms in the action, $\Delta I =
\frac{1}{2\pi} \int d^2x \sqrt{-g} \Phi [\sum a_n R^n]$, the change in
the entropy function is given by $\Delta F = - u \sum b_n v^{1-n}$,
where $b_n = (-2)^n a_n$. Then, the entropy~(\ref{entropy}) is modified as
\begin{equation}
  S_\mathrm{mod} = u_e [2 - \sum n b_n v_e^{1-n}]. \label{entropy:mod}
\end{equation}

The exact expression and implication of the black hole entropy
$S_\mathrm{BH}$ will be given through several interesting examples in
what follows.


\section{Two-dimensional effective heterotic string theory}
\label{sec:cghs}

First, we consider the two-dimensional gravity, which may come from
the compactification of the heterotic string theory, with the dilaton
field, $U(1)$ gauge field and the two-dimensional cosmological
constant. The action is given by
\begin{equation}
   I = \frac{1}{2\pi} \int d^2 x \sqrt{-g} e^{-2\phi} \left[
     R + 4 (\nabla \phi)^2 + 4 \lambda^2 - \frac14 F^2 \right]. \label{action:cghs}
\end{equation}
This theory was used in the Callan-Giddings-Harvey-Strominger(CGHS)
model~\cite{cghs} to study the quantum nature of the black holes. It
was known that the theory admits the static charged black hole
solutions of the form~\cite{mny}
\begin{eqnarray}
  ds^2 &=& -g(r) dt^2 + \frac{1}{g(r)} dr^2,  \nonumber \\
  \phi &=& -\lambda r, \label{metric:cghs} \\
  F_{rt} &=& \sqrt2 Q e^{-2\lambda r}, \nonumber
\end{eqnarray}
where $g(r)=1 - (M/\lambda) e^{-2\lambda r} + (Q^2/4\lambda^2)
e^{-4\lambda r}$.
In the extremal limit, $M=Q$, the metric function is reduced to $g(r) =
[1-(Q/2\lambda) e^{-2\lambda r}]^2$ and the near horizon geometry in
the leading order is clearly given by $AdS_2$ type. The near horizon
solutions which has $SO(2,1)$ symmetry can be written as
eq.~(\ref{sols:NH}) with $\Phi=e^{-2\phi}$. In this case
the Lagrangian density function is evaluated as
\begin{eqnarray}
  f(v,u,\varepsilon) &=& \frac{1}{2\pi} \sqrt{-g} e^{-2\phi} \left[
     R + 4 (\nabla \phi)^2 + 4\lambda^2 - \frac14 F^2 \right] \nonumber \\
  &=& \frac{v u}{2\pi} \left[ -\frac2v + 4\lambda^2 + \frac{1}{2v^2} \varepsilon^2 \right],
     \label{L:density:cghs}
\end{eqnarray}
where the electric charge is given by
\begin{equation}
q = \frac{\partial f}{\partial \varepsilon} = \frac{u}{2\pi v}
  \varepsilon .  \label{charge:cghs}
\end{equation}
Then, the entropy function is written as
\begin{equation}
  F = 2\pi [ q\varepsilon - f] = v u \left[ \frac2v - 4\lambda^2 + \frac{2\pi^2q^2}{u^2} \right],
  \label{ent:fnc:cghs}
\end{equation}
and extremizing eq.~(\ref{ent:fnc:cghs}) with respect to $v$ and $u$,
\begin{eqnarray}
  \left. \frac{\partial F}{\partial v} \right|_{(u_e,v_e)} &=& \left[ -4\lambda^2 u_e +
    \frac{2\pi^2q^2}{u_e} \right] = 0,  \label{v:eqn:cghs} \\
  \left. \frac{\partial F}{\partial u} \right|_{(u_e,v_e)} &=& \left[ 2 -4\lambda^2 v_e -
    \frac{2\pi^2v_eq^2}{u_e^2} \right] = 0.  \label{u:eqn:cghs}
\end{eqnarray}
yields solutions $u_e^2 = {\pi^2q^2}/{2\lambda^2}$ and $v_e = {1}/{4\lambda^2}$
in terms of the black hole charge, $q$ and the cosmological constant,
$\lambda$. In fact, these are consistent with the near horizon limit
of the extremal black holes in eq.~(\ref{metric:cghs}) with $M=Q$ as it
should be. Note that, in taking the near horizon limit we should make
replacements,
 $ 4\lambda^2 (r-r_\mathrm{H})\rightarrow r$ and $q =  \frac{Q}{\sqrt2\pi}$.

Plugging these values $u_e$ and $v_e$ into the entropy function, one finds the
black hole entropy 
\begin{equation}
  S_\mathrm{BH} = 2u_e = \frac{\sqrt{2}\pi |q|}{\lambda}, \label{entropy:cghs}
\end{equation}
which agrees to the result in ref.~\cite{np}.

Now, considering generic higher-derivative correction terms,
the extremizing equations~(\ref{v:eqn:cghs}) and~(\ref{u:eqn:cghs}) in
the CGHS-Maxwell model are modified as
\begin{eqnarray}
  \left. \frac{\partial F}{\partial v} \right|_{(u_e,v_e)} &=& \left[ -4\lambda^2 u_e +
    \frac{4\pi^2q^2}{2u_e} - u_e \sum_{n=2} b_n (1-n) v_e^{-n} \right] = 0,  \label{v:eqn:mod} \\
  \left. \frac{\partial F}{\partial u} \right|_{(u_e,v_e)} &=& \left[ 2 -4\lambda^2 v_e -
    \frac{4\pi^2v_eq^2}{2u_e^2} - \sum_{n=2} b_n v_e^{1-n} \right] = 0.  \label{u:eqn:mod}
\end{eqnarray}
Using these equations, the entropy~(\ref{entropy:mod}) can be written
in terms of $q$, $S_\mathrm{mod} = 4\pi^2 v_e q^2/u_e$. In this
two-dimensional set-up, the only dimensionful geometric quantity is
$\lambda$. Checking the dimension of total action $I+\Delta I$, we see
that the coefficient $a_n$ has the dimension of order
$\lambda^{2(1-n)}$, and it is straightforward to check that $u$ is of
order $q/\lambda$ and $v$ is of order $\lambda^{-2}$ from eqs.~(\ref{v:eqn:mod})
and (\ref{u:eqn:mod}). Then, it is easily seen that the modified
entropy is of order $q/\lambda$, which yields that higher-derivative
correction terms only modify the factor of the original
entropy~(\ref{entropy:cghs}) and can be written generically as
$S_\mathrm{BH} = a\frac{|q|}{\lambda} $, where $a$ is the pure
numerical factor depending on the higher derivative terms.

For example, one may consider four-dimensional Gauss-Bonnet terms,
which appear in the low energy effective theory of heterotic strings,
of the form
\begin{equation}
\Delta I =   \frac{1}{16\pi}\int d^4 x \sqrt{-g_{(4)}}e^{-2\phi_{(4)}} \{ R_{(4)\mu\nu\rho\sigma}R_{(4)}^{\mu\nu\rho\sigma}-4 R_{(4)\mu\nu}R_{(4)}^{\mu\nu} +R_{(4)}^2\}~.
\end{equation}
If the four-dimensional metric can be written as the direct product of
two-dimensional space and two-sphere, $S^2$, with constant radius, the
two-dimensional effective action after the dimensional reduction becomes
\begin{equation}
\Delta I = \frac{\alpha}{2\pi} \int d^2 x \sqrt{-g} e^{-2\phi} R 
\end{equation}
where the numerical constant $\alpha$ is related to the ratio of the
four-dimensional Newton constant and the volume of
two-sphere. Therefore the incorporation of the four-dimensional
Gauss-Bonnet terms change only the two-dimensional Newton constant,
and thus the $AdS$ geometry becomes the exact solution of the full
theory.  After including these higher-derivative terms, the 
black hole entropy gets modified and becomes 
\begin{equation}
  S_\mathrm{BH} = 2(1+\alpha)u_e = (1+\alpha)\frac{\sqrt{2}\pi |q|}{\lambda}. \label{mod-entropy:cghs}
\end{equation}

\section{JT model}
\label{sec:jt}

Another interesting model in two-dimensional gravity is  the
Jackiw-Teitelboim (JT) model~\cite{jt}.
In order to study the extremal charged black hole, one can include a
gauge field in the model with the following form of the action 
\begin{equation}
   I = \frac{1}{2\pi} \int d^2 x \sqrt{-g} e^{-2\phi} \left[
     R + 4 \lambda^2 - \frac14 e^{-4\phi} F^2 \right].  \label{action:jt}
\end{equation}
Again, one may regard this model as the dimensional reduction of a
higher-dimensional gravity. Indeed the action can be obtained from the
Kaluza-Klein (KK) compactification along the circle direction of the
three-dimensional Einstein-Hilbert action with a cosmological
constant. It was known that the model also admits the static charged
solutions whose configurations are given by
\begin{eqnarray}
  ds^2 &=& -g(r) dt^2 + \frac{1}{g(r)} dr^2,  \nonumber \\
  e^{-2\phi} &=& \sqrt2 \lambda r,  \label{sols:jt} \\
  F_{rt} &=& \frac{Q}{2 \lambda^3 r^3},  \nonumber
\end{eqnarray}
where $g(r)=2\lambda^2 r^2 - M/\lambda + Q^2/8\lambda^4r^2$~\cite{ls}.
In the extremal limit, the metric function is reduced to $g(r) =
2\lambda^2r^2 [1-Q/4\lambda^3r^2]^2$. 
The near horizon solutions are those of $AdS_2$ with $SO(2,1)$
symmetry. Therefore one can study the black hole entropy in the
similar fashion.

The  entropy function, $F$, which is the Legendre transform of the
Lagrangian density function, $f$, in terms of the electric field becomes 
\begin{equation}
  F = 2\pi [ q\varepsilon - f] = v u \left[ \frac2v - 4\lambda^2 + \frac{2\pi^2q^2}{u^4} \right],
  \label{ent:fnc:jt}
\end{equation}
where the electric charge can be determined as $q = {\partial f}/{\partial \varepsilon} =
{u^3\varepsilon}/{2\pi v}$. One may see that with the action given
above, i.e. without higher derivative terms, the charge $q$ is related
to the parameter $Q$ of the black hole solutions as $q =
{Q}/{\sqrt2\pi}$.

Extremizing it with respect to $v$ and $u$,
\begin{eqnarray}
  \left. \frac{\partial F}{\partial v} \right|_{(u_e,v_e)} &=& \left[ -4\lambda^2 u_e +
    \frac{2\pi^2q^2}{u_e^3} \right] = 0,  \label{v:eqn:jt} \\
  \left. \frac{\partial F}{\partial u} \right|_{(u_e,v_e)} &=& \left[ 2 -4\lambda^2 v_e -
    \frac{6\pi^2v_eq^2}{u_e^4} \right] = 0.  \label{u:eqn:jt}
\end{eqnarray}
provides the extremizing solutions $u_e^4 = {\pi^2q^2}/{2\lambda^2}$
and $v_e = {1}/{8\lambda^2}$. By  plugging these back into the entropy
function, we obtain the black hole entropy as
\begin{equation}
  S_\mathrm{BH} = 2u_e = 2^{3/4}\sqrt{\frac{\pi|q|}{\lambda}},  \label{entropy:jt}
\end{equation}
which is exactly the same result as in
ref.~\cite{kr} except for the extra $1/\lambda$ factor.~\footnote{
  Note that the entropy~(\ref{entropy:jt}) is dimensionless in mass
  dimension, while that of ref.~\cite{kr} has length dimension.} One
may also consider the higher derivative corrections along the similar
fashion. It will give a numerical corrections to an overall factor as
discussed in the previous section.

\section{Doubly charged CGHS Model}
\label{sec:dc}


Finally, we consider the doubly charged CGHS model which can be
obtained by the dimensional reduction from the ten-dimensional IIB
supergravity action coupled to the two-form Ramond-Ramond gauge
field. One may consider the type IIB black holes on $M_5 \times S^1 \times T^4$,
where D5-brane wraps the five torus $S^1 \times T^4$ $Q_5$ times, D
strings wrap the circle $S^1$ $Q_2$ times, and a 
KK momentum $Q$ runs along the circle $S^1$~\cite{klp:thermo,klp}.
Compactifying $M_5$ to $M_2 \times S^3$, the reduced two-dimensional
action is given by
\begin{equation}
  I = \frac{1}{2\pi} \int d^2 x \sqrt{-g} e^{-2\phi} \left[
    R + 4 (\nabla \phi)^2 + 4 \lambda^2 - (\nabla\psi)^2 - \frac14
    (\nabla\psi_1)^2 - \frac14 e^{-\psi_1} F_2^2 - \frac14 e^{\psi_1}
    F^2 \right], \label{action:dc}
\end{equation}
where the cosmological constant $\lambda$ is related to the radius of
$S^3$, $\lambda = e^{\psi_2}$, which is assumed to be constant.
The field strength $F=dA$ has the KK momentum as its
charge and the field strength $F_{2\alpha\beta}=H_{\alpha\beta x^5}$
comes from the string wrapping along the circle $S^1$.

The static solutions are given by
\begin{eqnarray}
  ds^2 &=& - \frac{\beta^2(1-r_0^2/r^2)}{(1+r_1^2/r^2)(1+r_2^2/r^2)}
    dt^2 + \frac{dr^2}{\lambda^2 r^2 (1-r_0^2/r^2)},  \label{met:dc} \\
  e^{-2\phi} &=& \frac{\sqrt{(r^2 + r_1^2)(r^2 + r_2^2)}}{\lambda},  \label{dilaton:dc} \\
  F_{rt} &=& \frac{2rQ\beta}{(r^2+r_2^2)^2},  \label{EM:dc} \\
  F_{2rt} &=& \frac{2rQ_2\beta}{(r^2+r_1^2)^2},  \label{EM2:dc}
\end{eqnarray}
where  $r_1^2 = \sqrt{Q_2^2 +
r_0^4/4} - r_0^2/2$ and $r_2^2 = \sqrt{Q^2 + r_0^4/4} -
r_0^2/2$ are constant, and $r_0\to0$ in the extremal
limit~\cite{klp}. The near horizon geometry of the extremal charged
black hole solution becomes $AdS_2$ and the field configurations
become symmetric under the $SO(2,1)$. Generically, the field
configurations which are consistent with $SO(2,1)$ are given by
eq.~(\ref{sols:NH}) for the metric, dilaton, and one gauge field
along with the following field configurations for the other two
scalars and extra gauge field as
\begin{eqnarray}
   \psi &=& u_\psi, \nonumber\\
  \psi_1 &=& \ln u_{\psi_1}, \label{psi:NH:dc} \\
  F_{2\tilde{r}t} &=& \varepsilon_2. \nonumber
\end{eqnarray}
The
Lagrangian density function is, then, given by
\begin{equation}
  f(v,u,\varepsilon) = \frac{v u}{2\pi} \left[ -\frac2v + 4\lambda^2 +
  \frac{u_{\psi_1}}{2v^2} \varepsilon^2 + \frac{1}{2v^2u_{\psi_1}}
  \varepsilon_2^2 \right],  \label{L:density:dc}
\end{equation}
from which one can determine the electric charges of the black hole as 
\begin{eqnarray}
  q = \frac{uu_{\psi_1}}{2\pi v} \varepsilon,  \qquad
  q_2 = \frac{u}{2\pi vu_{\psi_1}} \varepsilon_2.  \label{charge2:dc}
\end{eqnarray}
The entropy function becomes
\begin{eqnarray}
  F = v u \left[ \frac2v - 4\lambda^2 +
    \frac{2\pi^2q^2}{u^2u_{\psi_1}} + \frac{2\pi^2u_{\psi_1}q_2^2}{u^2}
    \right],  \label{ent:fnc:dc}
\end{eqnarray}
whose extremum is determined by 
\begin{eqnarray}
  \left. \frac{\partial F}{\partial v} \right|_{(u_e,v_e,u_{e\psi_1})} &=& \left[ -4\lambda^2 u_e +
    \frac{2\pi^2q^2}{u_eu_{e\psi_1}} + \frac{2\pi^2u_{e\psi_1}q_2^2}{u_e} \right] = 0,
    \label{v:eqn:dc} \\
  \left. \frac{\partial F}{\partial u} \right|_{(u_e,v_e,u_{e\psi_1})} &=& \left[ 2 -4\lambda^2 v_e -
    \frac{2\pi^2v_eq^2}{u_e^2u_{e\psi_1}} - \frac{2\pi^2v_eu_{e\psi_1}q_2^2}{u_e^2}
    \right] = 0,  \label{u:eqn:dc} \\
  \left. \frac{\partial F}{\partial u_{\psi_1}} \right|_{(u_e,v_e,u_{e\psi_1})} &=& \left( \frac{4\pi^2v_e}{2u_e}
    \right) \left[ - \left(\frac{q^2}{u_{e\psi_1}^2} \right) + q_2^2
    \right] = 0.  \label{upsi:eqn:dc}
\end{eqnarray}
The extremizing solutions are
\begin{equation}
  u_e^2 = \frac{\pi^2|qq_2|}{\lambda^2}, \qquad v_e = \frac{1}{4\lambda^2},
  \qquad u_{e\psi_1}^2 = \frac{q^2}{q_2^2}.  \label{ext:sol:dc}
\end{equation}
Note that  with the help of the
extremizing solutions~(\ref{ext:sol:dc}), one can determine the
relations between the black hole charges ($q$, $q_2$) and the
parameters of the black hole solutions ($Q$, $Q_2$) as $q = Q/\pi$ and
$q_2 = Q_2/\pi$.
Plugging the extremum values~(\ref{ext:sol:dc}) into the entropy
function finally yields the black hole entropy as
\begin{equation}
  S_\mathrm{BH} = 2u_e = \frac{2\pi \sqrt{|qq_2|}}{\lambda},  \label{entropy:dc}
\end{equation}
which is exactly the same result as in
ref.~\cite{klp:thermo}.  It is straightforward to include the higher
derivative terms and find out the corrections to the entropy.
There are another extremal black hole solutions in a non-perturbatively 
defined string theory of the two-dimensional type 0 string, which has the matrix model 
description~\cite{gtt,str:mtx}. They also have the near horizon
geometry of $AdS_2$, but we will not consider them in this paper.

\section{Discussions}
\label{sec:discussion}

For black holes with supersymmetry, a successful statistical
understanding of the thermodynamic Bekenstein-Hawking entropy has been
achieved by counting microstates~\cite{sv}. This resulted from the BPS
property of supersymmetric black holes - the BPS states can be
computed at weak coupling regime and then analytically continued to
the strong coupling regime. On the other hand, the attractor mechanism that
the macroscopic entropy is independent of the asymptotic values of the
moduli is closely related to the near horizon geometry of extremal
black holes. Both significant properties simplified the issue of
evaluating the entropy of supersymmetric black holes and offered the
agreement between the thermodynamic and the statistical entropies.
Motivated by the fact that there are many indications of this
agreement for non-supersymmetric black holes, it has been conjectured
that the same agreement still work and the origin does not rely on
supersymmetry but the attractor mechanism of all extremal black holes
whether or not they are supersymmetric~\cite{dst}.
More precisely, the thermodynamic Bekenstein-Hawking entropy of all
extremal black holes within string theory, in particular, showing a near-horizon
geometry of $AdS_2 \times S^{D-2}$ exactly matches with the
statistical microscopic entropy at weak coupling regime. 
As seen in ref.~\cite{dst}, there are a variety of examples of
non-supersymmetric black holes that support
the surmised agreement between the macroscopic and the
microscopic entropies for all extremal black holes.

In this paper, using the Sen's entropy function, we found the
universal behavior of the black hole entropy in the two-dimensional
gravity. We also considered the higher derivative corrections and
found the universal behavior remains to hold.  In all cases we
considered, the extremal black hole entropy only depends on the field
configuration at the horizon, regardless of its asymptotic values of
the fields. In this sense, these examples also show the attractor
mechanism of the black hole entropy~\cite{fks,str,fk}.
Hence, our
studies on the entropy function method of two-dimensional black holes
might be another good examples of finding the entropy of
non-supersymmetric black holes, which robustly upholds the underlying
conjecture.

One motivation to study the black hole entropy in the context of the
two-dimensional gravity is to probe the quantum nature of gravity. In
the two-dimensional gravity, the quantum corrections of the scalar
field typically include Polyakov-Liouville (PL)
action~\cite{rst,bpp}. In fact, we considered this issue and found that 
the entropy function method may not be applied to two-dimensional
quantum-corrected model because the nonlocality in PL term produces an
inconsistent extremal equations. It seems to imply that the near
horizon geometry is no longer $AdS_2$ type after the quantum
corrections. Even if the near horizon solution is reduced to
$AdS_2$, there remains another problem. Localizing PL action, by introducing an
auxiliary field, say $\Psi$, requires that $\Box \Psi = R$. In the
near horizon limit, the equation is written as
$\partial_{\tilde{r}} \tilde{r}^2 \partial_{\tilde{r}} \Psi = -2$, and
the solution, $\Psi = -2 \ln \tilde{r} + c_1/\tilde{r} + c_2$, is
divergent at the horizon, which violates the assumption that the
scalar field near horizon has a finite constant value. Further
studies on this issue is needed.

\vspace{1cm}


{\bf Acknowledgments}

The work of S.~Hyun was partially supported by
the Basic Research Program of the Korea Science and Engineering
Foundation under grant number R01-2004-000-10651-0.
S.~Hyun and W.~Kim were supported by the Science Research Center Program of the
Korea Science and Engineering Foundation through the Center for
Quantum Spacetime \textbf{(CQUeST)} of Sogang University with grant
number R11 - 2005 - 021. J.~J.~Oh was supported by the Brain Korea 21(BK21) 
project funded by the Ministry of Education and Human Resources of Korea Government.
E.~J.~Son was supported by the Korea Research Foundation Grant funded
by Korea Government(MOEHRD, Basic Reasearch Promotion Fund)
(KRF-2005-070-C00030).



\begin{thebibliography}{99}
\bibitem{cghs}
  C.~G.~Callan, S.~B.~Giddings, J.~A.~Harvey, and A.~Strominger,
  Phys.\ Rev.\ D {\bf 45}, 1005 (1992)
  [hep-th/9111056].
\bibitem{jt}
  R.~Jackiw,
  Nucl.\ Phys.\ B {\bf 252}, 343 (1985);
  C.~Teitelboim,
  Phys.\ Lett.\ B {\bf 126}, 41 (1983).
\bibitem{np}
  C.~R.~Nappi and A.~Pasquinucci,
  Mod.\ Phys.\ Lett.\ A {\bf 7}, 3337 (1992)
  [gr-qc/9208002].
\bibitem{kr}
  A.~Kumar and K.~Ray,
  Phys.\ Lett.\ B {\bf 351}, 431 (1995)
  [hep-th/9410068].
\bibitem{cm}
  M.~Cadoni and S.~Mignemi,
  Phys.\ Rev.\ D {\bf 59}, 081501 (1999)
  [hep-th/9810251].
\bibitem{cs}
  M.~Cadoni and N.~Serra,
  Phys.\ Rev.\ D {\bf 70}, 126003 (2004)
  [hep-th/0406153].
\bibitem{klp:thermo}
  Y.~Kiem, C.~Y.~Lee, and D.~Park,
  Phys.\ Rev.\ D {\bf 58}, 125002 (1998)
  [hep-th/9806182].
\bibitem{sw}
  D.~Sudarsky and R.~M.~Wald,
  Phys.\ Rev.\ D {\bf 46}, 1453 (1992);
  J.~D.~Brown and J.~W.~York,
  Phys.\ Rev.\ D {\bf 47}, 1407 (1993);
  J.~D.~Brown and J.~W.~York,
  Phys.\ Rev.\ D {\bf 66}, 010001 (2002)
  [gr-qc/9209014].
\bibitem{wald}
  R.~M.~Wald,
  Phys.\ Rev.\ D {\bf 48}, 3427 (1993)
  [gr-qc/9307038];
  T.~Jacobson, G.~Kang, and R.~C.~Myers,
  Phys.\ Rev.\ D {\bf 66}, 010001 (2002)
  [gr-qc/9312023];
  V.~Iyer and R.~M.~Wald,
  Phys.\ Rev.\ D {\bf 50}, 846 (1994)
  [gr-qc/9403028];
  T.~Jacobson, G.~Kang, and R.~C.~Myers,
  Phys.\ Rev.\ D {\bf 66}, 010001 (2002)
  [gr-qc/9502009].
\bibitem{sen}
  A.~Sen,
  JHEP {\bf 0509}, 038 (2005)
  [hep-th/0506177].
\bibitem{mny}
  M.~D.~McGuigan, C.~R.~Nappi, and S.~A.~Yost,
  Nucl.\ Phys.\ B {\bf 375}, 421 (1992)
  [hep-th/9111038].
\bibitem{ls}
  D.~A.~Lowe and A.~Strominger,
  Phys.\ Rev.\ D {\bf 66}, 010001 (2002)
  [hep-th/9403186].
\bibitem{klp}
  Y.~Kiem, C.~Y.~Lee, and D.~Park,
  Phys.\ Rev.\  D {\bf 57}, 2381 (1998)
  [hep-th/9705065].
\bibitem{gtt}
  S.~Gukov, T.~Takayanagi, and N.~Toumbas,
  JHEP {\bf 0403}, 017 (2004)
  [hep-th/0312208].
\bibitem{str:mtx}
  A.~Strominger,
  JHEP {\bf 0403}, 066 (2004)
  [hep-th/0312194].
\bibitem{sv}
  A.~Strominger and C.~Vafa,
  Phys.\ Lett.\  B {\bf 379}, 99 (1996)
  [hep-th/9601029].
\bibitem{dst}
  A.~Dabholkar, A.~Sen, and S.~P.~Trivedi,
  JHEP {\bf 0701}, 096 (2007)
  [hep-th/0611143].
\bibitem{fks}
  S.~Ferrara, R.~Kallosh, and A.~Strominger,
  Phys.\ Rev.\  D {\bf 52}, 5412 (1995)
  [hep-th/9508072].
\bibitem{str}
  A.~Strominger,
  Phys.\ Lett.\  B {\bf 383}, 39 (1996)
  [hep-th/9602111].
\bibitem{fk}
  S.~Ferrara and R.~Kallosh,
  Phys.\ Rev.\  D {\bf 54}, 1514 (1996)
  [hep-th/9602136].
\bibitem{rst}
  J.~G.~Russo, L.~Susskind, and L.~Thorlacius,
  Phys.\ Rev.\ D {\bf 46}, 3444 (1992)
  [hep-th/9206070].
\bibitem{bpp}
  S.~Bose, L.~Parker, and Y.~Peleg,
  Phys.\ Rev.\ Lett.\  {\bf 76}, 861 (1996)
  [gr-qc/9508027].
\end{thebibliography}
\end{document}